\documentclass[english,pra,showpacs,showkeys,tightenlines,secnumarabic,11pt]{revtex4}
\usepackage[T1]{fontenc}
\usepackage[latin1]{inputenc}
\usepackage{amsmath}
\usepackage{graphicx}
\usepackage{amssymb}
\usepackage{epsfig}
\usepackage{graphics}
\usepackage[mathscr]{euscript}
\usepackage{psfrag}
\usepackage{pstricks}
\usepackage{pst-node}
\usepackage{babel}

\newcommand{\esp}[1]{\, e^{\,\,\textstyle {#1}}}

\begin{document}
\title{Incoherence and Multiple Parton Interactions}
\author{G. Calucci}
\email{giorgio.calucci@ts.infn.it}
\author{D. Treleani}
\email{daniele.treleani@ts.infn.it} \affiliation{ Dipartimento di
Fisica Teorica dell'Universit\`a di Trieste and INFN, Sezione di
Trieste,\\ Strada Costiera 11, Miramare-Grignano, I-34151 Trieste,
Italy.}
\begin{abstract}
At the LHC Multiple Parton Interactions will represent an
important feature of the minimum bias and of the underlying event
and will give important contributions in many channels of
interest for the search of new physics. Different numbers of multiple collision may contribute to the production of a given final state and one should expect important interference effects in the regime where different contributions have similar rates. We show, on the contrary, that, once multiple parton interactions are identified by their different topologies, terms with different numbers of multiple parton interactions do not interfere in the final cross section.
\end{abstract}



 \maketitle

\section{Introduction}

The growing flux of partons at high energy will increase considerably the chances to have inelastic events where more than a single pair of partons interact with large momentum exchange at the LHC\cite{HERA-LHC}\cite{Perugia}\cite{Kulesza:1999zh}\cite{Acosta:2004wqa}\cite{Acosta:2006bp}\cite{Hussein:2007gj}\cite{Maina:2009vx}\cite{Domdey:2009bg}. The phenomenon is originated by the increasingly large flux of partons at small fractional momenta. Once the final state is fixed the parton flux is maximized in the channel where the hard component of the interaction is maximally disconnected\cite{Paver:1984ux}. In the resulting picture of the process, the dominant contribution at high energy is hence given by a set of Multiple Parton Interactions (MPI) where different pairs of partons collide independently in different points inside the overlap volume of the two interacting hadrons\cite{Sjostrand:1987su}\cite{Ametller:1987ru}\cite{Rogers:2008ua}. On the other hand, although the contribution with the largest number of initial state partons dominates at large hadron-hadron center of mass energes, a given final state may be generated by various competing processes, characterized by different numbers of partonic collisions\cite{Maina:2009vx}. Interactions with different numbers of partonic collisions populate in fact the final state phase space in a different way and one may always find kinematical regions where terms with different numbers of collisions give similar contributions to the cross section. In those kinematical regions, important interference effects between different production mechanisms should be expected. The problem of interferences between terms with different numbers of collisions was, on the other hand, never discussed in the literature, while all theoretical estimates have always assumed incoherence between contributions with different numbers of parton collisions, obtaining results not in contradiction with the available experimental evidence\cite{Akesson:1986iv}\cite{Abe:1997bp}\cite{Abe:1997xk}\cite{Abazov:2009wy}.

The purpose of the present note is to gain some understanding of the problem by looking at the kinematics of the different terms. After
reminding the reader of the kinematical argument, which leads to the geometrical
picture of MPI processes, we will analyze an
interference diagram. The comparison between diagonal and off-diagonal terms in the cross section will allow to draw some general conclusions.

\section{MPI diagonal scattering diagram}

In Fig.1 we show the cut diagram representing the contribution to the forward amplitude of a process with $n$ partonic collisions.

\begin{figure}[htp]
\centering
\includegraphics[width=13cm]{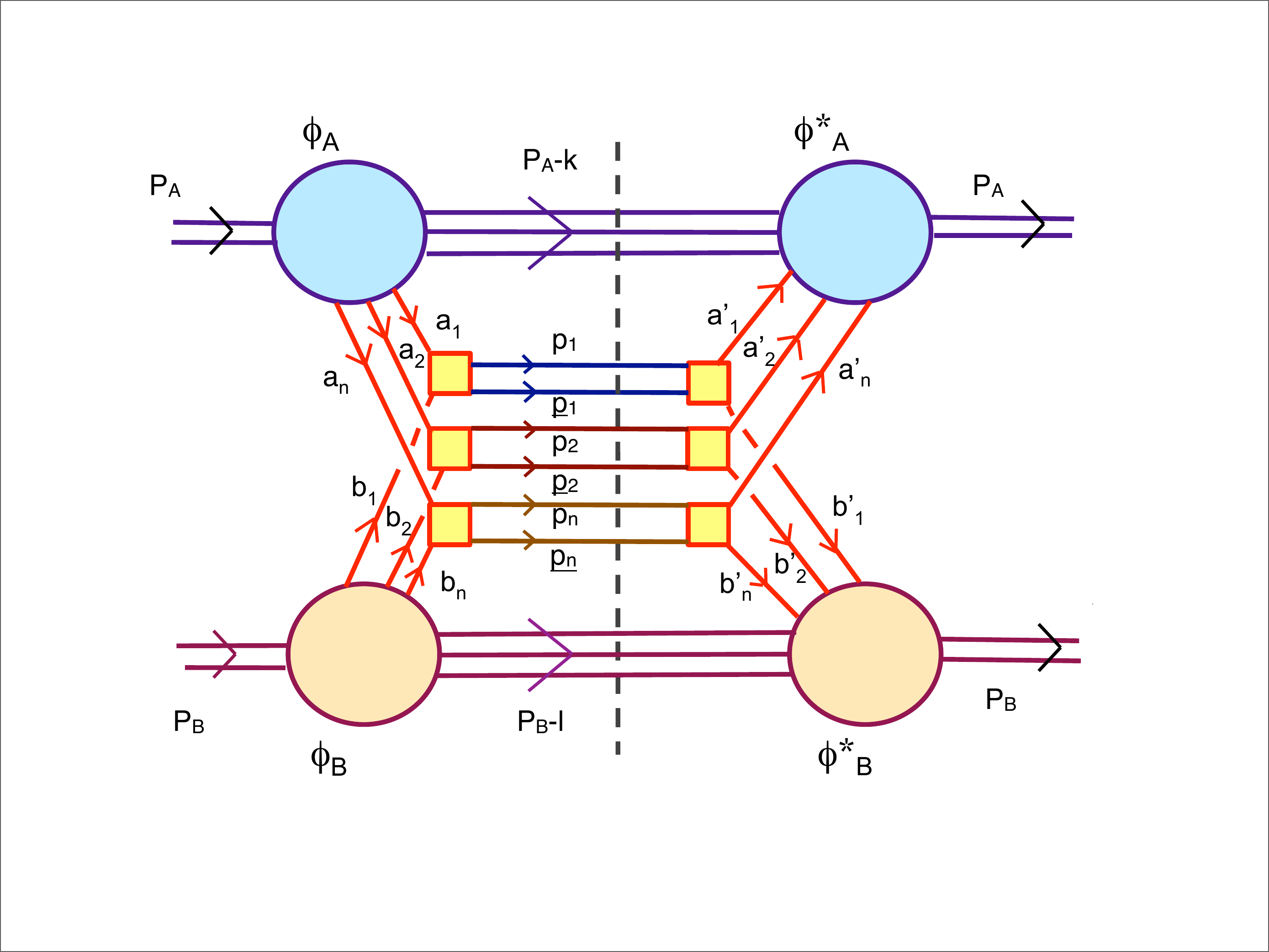}
\vspace{0cm}
\caption{Unitarity diagram for the multi parton scattering cross section}
\label{fig:MPI}
\end{figure}

\noindent
To study the kinematics we simplify the problem by limiting our discussion to the scalar case\cite{Paver:1982yp}. We consider moreover all partons as identical particles. The soft vertices $\phi$ are assumed to be characterized by a non-perturbative scale of the order of the hadron radius $R$, which represents the (energy independent) scale of the transverse momenta and virtualities of the attached lines. To be definite, we consider the specific multi-parton interaction process where each elementary interaction $T_i$, represented by the squares in the figure, generates two large $p_t$ partons with momenta $p_i$ and $\bar p_i$.

\noindent
Momentum conservation in the vertices limits the number of four dimensional variables to $3n-1$. Defining

\begin{eqnarray}
\delta_i=\frac{1}{2}(a_i-a_{i+1}),\qquad \delta_i'=\frac{1}{2}(a_i'-a_{i+1}'),\qquad P_i=p_i+\bar p_i,
\end{eqnarray}

\noindent
in such a way that $a_i+b_i=P_i=a_1'+b_i'$, one may choose as independent variables:

\begin{eqnarray}
\begin{cases}
P_i,&n-\text{variables},\\
\delta_i,&(n-1)-\text{variables},\\
\delta_i',&(n-1)-\text{variables},\\
k\quad\text{or}\quad l,
\end{cases}
\end{eqnarray}

\noindent
since the overall momentum conservation is $\sum_iP_i+k+l=P_A+P_B$. Sometimes the auxiliary dependent variables

\begin{eqnarray}
\bar\delta_i=\frac{1}{2}(b_i-b_{i+1})=\frac{1}{2}(P_i-P_{i+1})-\delta_i
\end{eqnarray}

\noindent
will be used.

The four momentum variables will be represented in the light-cone form and the longitudinal and transverse components will be studied separately. The vertices $\phi$ are "soft" and represent the non-perturbative partonic content of the hadron. In the hadron c.m. one has

\begin{eqnarray}
&&a_{\perp}\lesssim\frac{1}{R}\nonumber\\
&&a_z\lesssim\frac{1}{R}\\
&&a^2\lesssim\frac{1}{R^2}\to a_0\lesssim\frac{1}{R},\nonumber
\end{eqnarray}

\noindent
where $R$ is the hadron radius. By performing a boost to reach the c.m. frame of the two interacting hadrons, one obtains

\begin{eqnarray}
\begin{cases}
a_{\perp},\ b_{\perp}\lesssim\frac{1}{R}\\
a_{+},\ b_{-}\lesssim\frac{\sqrt {\cal S}}{2M}\frac{1}{R}\\
a_{-},\ b_{+}\lesssim\frac{2M}{\sqrt {\cal S}}\frac{1}{R}\\
P_i^2=(a_i+b_i)^2\lesssim\frac{{\cal S}}{4M^2}\frac{1}{R^2},
\end{cases}
\end{eqnarray}

\noindent
where ${\cal S}=(P_A+P_B)^2$ is the hadron-hadron c.m. energy and $M$ is a scale of the order of the hadron mass. When looking at $\delta_{i-}$ and $\bar\delta_{i+}$ one hence finds that both variables become smaller and smaller at large c.m. energies:

\begin{eqnarray}
\begin{cases}
&\delta_{i-}\lesssim\frac{2M}{\sqrt {\cal S}}\frac{1}{R}\\
&\bar\delta_{i+}\lesssim\frac{2M}{\sqrt {\cal S}}\frac{1}{R}.
\end{cases}
\end{eqnarray}

\noindent
The variables $\delta_{i-}$ are thus relevant only for the vertex $\phi_A$ and for the propagators of the lines with momenta $a_i$, since the kinematical range of $a_{i-}$ is of  ${\cal O}(2M/R\sqrt {\cal S})$. At the leading order, one needs in fact to take into account only of the kinematical variables which grow as $\sqrt {\cal S}$ in the hard interaction vertices, while the lower vertex $\phi_B$ and the propagators of the lines with momenta $b_i$ will remain practically constant for variations of $\delta_-$ of ${\cal O}(2M/R\sqrt {\cal S})$, as the kinematical range of $b_{i-}$ is of ${\cal O}(\sqrt {\cal S})$.  Conversely the variables $\bar\delta_{i+}$ are relevant only for the vertex $\phi_B$ and for the propagators of the lines with momenta $b_i$, where all "+" components have a kinematical range of  ${\cal O}(2M/R\sqrt {\cal S})$. Similar considerations hold for the variables $a_i'$, $b_i'$, $\delta_{i-}'$, $\bar\delta_{i+}'$ and for the vertices $\phi_A'$, $\phi_B'$. One may hence define:

\begin{eqnarray}
&&\psi_A\bigl(a_{1+}\dots a_{n+};a_{1\perp}\dots a_{n\perp};k_-;j  \bigr)\equiv\int\frac{\phi_A(a_1\dots a_n,k;j)}{\prod_i^n a_i^2}\prod_i^{n-1} \frac{d\delta_{i-}}{2\pi}\Big|_{\bar\delta_{i+}=0}\nonumber\\
&&\psi_B\bigl(b_{1-}\dots b_{n-};b_{1\perp}\dots b_{n\perp};l_+;j'\bigr)\equiv\int\frac{\phi_B(b_1\dots b_n,l;j')}{\prod_i^n b_i^2}\prod_i^{n-1} \frac{d\delta_{i+}}{2\pi}\Big|_{\bar\delta_{i-}=0}
\end{eqnarray}

\noindent
where $\ k_+,\ k_{\perp}$ are given in terms of $a_{i+},\ a_{i\perp}$ and $\ l_-,\ l_{\perp}$ are given in terms of $b_{i-},\ b_{i\perp}$, while the value of $k_-$ is determined by the values of $k_+,\ k_{\perp}$ and by the value of the invariant mass of the remnants of the hadron $A$, $(P_A-k)^2$. $l_+$ is similarly determined by $l_-,\ l_{\perp}$ and by the value of the invariant mass of the remnants of the hadron $B$, $(P_B-l)^2$. All other variables which characterize the remnants of $A$ and $B$ are labeled by the indices $j$ and $j'$ respectively. One has:

\begin{eqnarray}
&&a_i+b_i=P_i=a_i'+b_i'\nonumber\\
&&a_{i+}+b_{i+}=a_{i+}'+b_{i+}'\simeq a_{i+}\simeq a_{i+}'=P_{i+}\\
&&a_{i-}+b_{i-}=a_{i-}'+b_{i-}'\simeq b_{i-}\simeq b_{i-}'=P_{i-}.\nonumber
\end{eqnarray}

\noindent
The "+" and "-" components $\frac{a_{i+}}{\sqrt {\cal S}},\ \frac{b_{i-}}{\sqrt {\cal S}}$ are the "+" or "-" fractional momenta $x_i^A,\ x_i^B$ and are given by the final state observable quantities $P_{i+}$ and $P_{i-}$. All longitudinal variables are hence either integrated or determined by the final state observables. In particular, taking into account only the terms which grow with $\sqrt {\cal S}$, one has that $x_i^A=x_i'^A$,  $x_i^B=x_i'^B$.

As we are particularly interested in the structure of the interaction in transverse plane, we express the elementary interaction amplitude as a two-dimensional Fourier transform with respect to the relative transverse distance $r_{i}$:

\begin{eqnarray}
T(\lambda_i, t_{i\perp})=\frac{1}{2\pi}\int \esp{i t_{i\perp}\cdot r_i}\tilde T(\lambda_i,r_i)d^2 r_i
\end{eqnarray}

\noindent where  $t_i=\frac{1}{2}(a_i-p_i-b_i+\bar p_i)$ is the momentum transfer and all longitudinal variables are summarized by $\lambda_i$.

The multiparton cross section may hence be obtained by performing the $n$ two-dimensional integrations on $r_i,\ r'_i$ and in the following variables:

\begin{eqnarray}
\begin{cases}
&P_{i\perp}=a_{i{\perp}}+b_{i{\perp}}=a_{i{\perp}}'+b_{i{\perp}}',\qquad n\ \text{variables},\\
&q_{i{\perp}}=(a_{i{\perp}}-b_{i{\perp}})/2,\qquad n\ \text{variables},\\
&q_{i{\perp}}'=(a_{i{\perp}}'-b_{i{\perp}}')/2,\qquad n\ \text{variables}.
\end{cases}
\end{eqnarray}

\noindent
Momentum conservation imposes however a constraint on the integration variables. As the incoming hadrons have vanishing transverse momenta one has

\begin{eqnarray}
k_{\perp}+\sum_ia_{i{\perp}}=k_{\perp}+\sum_ia_{i{\perp}}'=0,\qquad l_{\perp}+\sum_ib_{i{\perp}}=l_{\perp}+\sum_ib_{i{\perp}}'=0,
\end{eqnarray}

\noindent
which imply

\begin{eqnarray}
\sum_i(q_{i{\perp}}-q_{i{\perp}}')=0
\end{eqnarray}

\noindent
in such a way that the cross section is given by $5n-1$ independent two-dimensional integrations.
The integrals on the transverse components are suitably performed by representing the $\psi$ functions as two-dimensional Fourier transforms with respect to the transverse parton coordinates  $s_i,\bar s_i, s_i', \bar s_i'$:

\begin{eqnarray}
\psi_A\bigl(x_1^A\dots x_n^A;\ a_{1\perp}&\dots& a_{n\perp};(P_A-k)^2;j\bigr)\nonumber\\&=&\int\tilde\psi_A\bigl(x_1^A\dots x_n^A;\ s_1\dots s_n;(P_A-k)^2;j\bigr)\prod_{i=1}^n\esp{i a_{i\perp}\cdot s_i}\frac{d^2s_i}{2\pi}\nonumber\\
\psi_A^*\bigl(x_1^A\dots x_n^A;\ a_{1\perp}'&\dots& a_{n\perp}';(P_A-k)^2;j\bigr)\nonumber\\&=&\int\tilde\psi_A^*\bigr(x_1^A\dots x_n^A;\ s_1'\dots s_n';(P_A-k)^2;j\bigr)\prod_{i=1}^n\esp{-i a_{i\perp}'\cdot s_i'}\frac{d^2s_i'}{2\pi}\nonumber\\
\psi_B\bigl(x_1^B\dots x_n^B;\ b_{1\perp}&\dots& b_{n\perp};(P_B-l)^2;j'\bigr)\nonumber\\&=&\int\tilde\psi_B(x_1^B\dots x_n^B;\ \bar s_1\dots\bar s_n;(P_B-l)^2;j'\bigr)\prod_{i=1}^n\esp{i b_{i\perp}\cdot \bar s_i}\frac{d^2\bar s_i}{2\pi}\nonumber\\
\psi_B^*\bigl(x_1^B\dots x_n^B;\ b_{1\perp}'&\dots& b_{n\perp}';(P_B-l)^2;j'\bigr)\nonumber\\&=&\int\tilde\psi_B^*\bigl(x_1^B\dots x_n^B;\ \bar s_1'\dots\bar s_n';(P_B-l)^2;j'\bigr)\prod_{i=1}^n\esp{-i b_{i\perp}'\cdot \bar s_i'}\frac{d^2\bar s_i'}{2\pi},\nonumber\\
\end{eqnarray}

\noindent
where the longitudinal components are expressed through the fractional momenta $x_i^A,\ x_i^B$. The constraint in Eq.(12) is imposed by introducing the further integration

\begin{eqnarray}
\frac{1}{(2\pi)^2}\int \esp{i\beta\sum(q_{i{\perp}}-q_{i{\perp}}')}d^2\beta
\end{eqnarray}

\noindent
where $\beta$ is the hadronic impact parameter. One may integrate on $\prod dP_{\perp},\ \prod dq_{\perp},\ \prod dq_{\perp}'$ by using the relations

\begin{eqnarray}
a_{i\perp}&=&\frac{1}{2}P_{i\perp}+q_{i\perp},\qquad b_{i\perp}=\frac{1}{2}P_{i\perp}-q_{i\perp}\nonumber\\
a_{i\perp}'&=&\frac{1}{2}P_{i\perp}+q_{i\perp}',\qquad b_{i\perp}'=\frac{1}{2}P_{i\perp}-q_{i\perp}'\\
t_{i\perp}&=& q_{i\perp}-\frac{1}{2}(p_{i\perp}-\bar p_{i\perp}),\qquad t_{i\perp}'=q_{i\perp}'-\frac{1}{2}(p_{i\perp}-\bar p_{i\perp}).\nonumber
\end{eqnarray}

\noindent
One obtains

\begin{eqnarray}
\int \frac{dP_{i\perp}}{(2\pi)^2}&\to& \delta\big((s_i+\bar s_i-s'_i-\bar s'_i)/2\big)\nonumber\\
\int \frac{dq_{i\perp}}{(2\pi)^2}&\to& \delta(s_i-\bar s_i+r_i+\beta)\nonumber\\
\int \frac{dq_{i\perp}'}{(2\pi)^2}&\to& \delta(s_i'-\bar s_i'+r_i'+\beta)
\end{eqnarray}

\noindent
which is equivalent to $\delta\bigr(s_i-s'_i-\frac{1}{2}(r_i'-r_i)\bigl)\delta (s_i-s'_i+\bar s_i-\bar s'_i)\delta(s_i-\bar s_i+r_i+\beta)$. The integrations on $r_i$ and $r_i'$ involve the factor

\begin{eqnarray}
\frac{1}{(2\pi)^2}\int \esp{i \frac{r_i-r_i'}{2}\cdot(t_{i\perp}-t_{i\perp}')}\tilde T(\lambda_i,r_i)\tilde T^*(\lambda_i,r_i')d^2 r_id^2 r_i'.
\end{eqnarray}

\noindent The difference $\frac{1}{2}(r_i'-r_i)$ hence represents a measure of the localization of the interaction. As it appears from eq.(9), typical values of $r_i$ and $r'_i$ are of the order of $1/t_{i\perp}$, which for large transverse momenta is equal to $1/p_{i\perp}$. When the momentum exchanged in the elementary interaction is large,  $r_i$ becomes much smaller as compared with the hadron size and it may be neglected, as well as $r'_i$, everywhere except in the expression in Eq.17. The dominant contribution at large c.m. energy and at large exchanged momenta is thus obtained by making the following replacements in the evaluation of the discontinuity of the diagram in Fig.1:

\begin{eqnarray}
\delta\bigr(s_i-s'_i-\frac{1}{2}(r_i'-r_i)\bigl)\delta (s_i-s'_i+\bar s_i-\bar s'_i)\delta(&s_i&-\bar s_i+r_i+\beta)\nonumber\\ \Longrightarrow&&\delta\bigr(s_i-s'_i\bigl)\delta\bigl(\bar s_i-\bar s'_i\bigl)\delta(s_i-\bar s_i+\beta)\nonumber\\
\frac{1}{(2\pi)^2}\int \esp{i \frac{r_i-r_i'}{2}\cdot(t_{i\perp}-t_{i\perp}')}\tilde T(\lambda_i,r_i)\tilde T^*(&\lambda_i&,r_i')d^2 r_id^2 r_i'\nonumber\\ \Longrightarrow &&|T(x_i,x'_i|p_i,\bar p_i)|^2,
\end{eqnarray}

\noindent where only the kinematical components which grow with ${\cal S}$ are taken into account in the evaluation of $T$.
One is hence left with the integrations on $\int d(P_A-k)_-d(P_B-l)_+d\beta\prod ds_i$, while the inclusive cross section depends on $|\tilde\psi_A|^2$ and $|\tilde\psi_B|^2$.  Explicitly:

\begin{eqnarray}
\sigma_n=\frac{1}{2{\cal S}n!}\int&&
\sum_j\big|\tilde\psi_A\bigl(x_1^A\dots x_n^A;s_1\dots s_n;(P_A-k)^2;j\bigr)\big|^2\nonumber\\
&\times&\sum_{j'}\big|\tilde\psi_B\bigl(x_1^B\dots x_n^B;s_1-\beta\dots s_n-\beta;(P_B-l)^2;j'\bigr)\big|^2\nonumber\\
&\times& d(P_A-k)_-d(P_B-l)_+\frac{d^2\beta}{(2\pi)^2}\Bigl[\frac{n}{2^{n-1}}\Bigr]^4\prod_i^n\frac{d^2s_i}{2(2\pi)^2}  \nonumber\\
& \times&\big|T(x_i^A,x_i^B|p_i,\bar p_i)\big|^2d\Phi_i \;\Bigl(\frac{\cal S}{2}\Bigr)^n \;dx_1^A\dots dx_n^A\;dx_1^B\dots dx_n^B
\end{eqnarray}

\noindent where $\Phi_i$ is the invariant adimensional final state phase space of the elementary interaction $a_i+b_i\to p_i+\bar p_i$ and all elementary interactions are considered as indistinguishable. The factors $2$, $\pi$ etc. originate from the Jacobian which leads to the variables in Eq.19.

After multiplying and dividing by the flux factors $2{\cal S}x_i^Ax_i^B$,  one may introduce the elementary partonic cross sections
\begin{eqnarray}
\hat\sigma(x_i^A,x_i^B;p_{cut})=\frac{1}{2{\cal S}x_i^Ax_i^B}\int_{p_{i\perp}>p_{cut}}\big|T(x_i^A,x_i^B|p_i,\bar p_i)\big|^2d\Phi_i
\end{eqnarray}

\noindent where $p_{cut}$ is a cutoff in the transverse momenta of final state partons, introduced to allow to compute the cross section in perturbative QCD. The multi-parton densities $\Gamma(x_i;s_i)$ are hence defined as

\begin{eqnarray}
\Gamma(x_1\dots x_n;s_1\dots s_n)=\frac{1}{(2\pi)^{n+1}}&&\frac{n^2}{4^{n-1}}\frac{{\cal S}^{n-1}\prod_i^nx_i}{\sqrt 2(1-\sum_i^nx_i)}\nonumber\\\times&&\int\sum_j \big|\tilde\psi\bigl(x_1\dots x_n;s_1\dots s_n;(P-k)^2;j\bigr)\big|^2 d(P-k)^2
\end{eqnarray}

\noindent and the cross section is finally expressed by

\begin{eqnarray}
\sigma_n=\frac{1}{n!}\int&&\Gamma_A(x_1^A\dots x_n^A;s_1\dots s_n)\nonumber\\
&\times&\prod_i\hat\sigma(x_i^A,x_i^B;p_{cut})\Gamma_B(x_1^B\dots x_n^B;s_1-\beta\dots s_n-\beta)d^2\beta d^2s_idx_i^Adx_i^B,
\end{eqnarray}

\noindent
which represents the superposition of $n$ elementary collisions localized in regions with transverse size of the order of $1/p_{cut}$, much smaller as compared with the hadron size, and with the mean value of the transverse coordinates $s_i$.

\section{Interference terms}

There are two different ways of producing interference terms. A possibility is to have a different number of hard collisions on the left and on the right hand side of the cut. Another possibility is to have the same number of hard collisions on both sides of the cut, in which case interferences are produced by reshuffling the final states of the hard collision terms.

We will first consider a case of interference between a term with $n$ and a term with $n-1$ collisions, the corresponding unitarity diagram is shown in Fig.2.

\begin{figure}[htp]
\centering
\includegraphics[width=13cm]{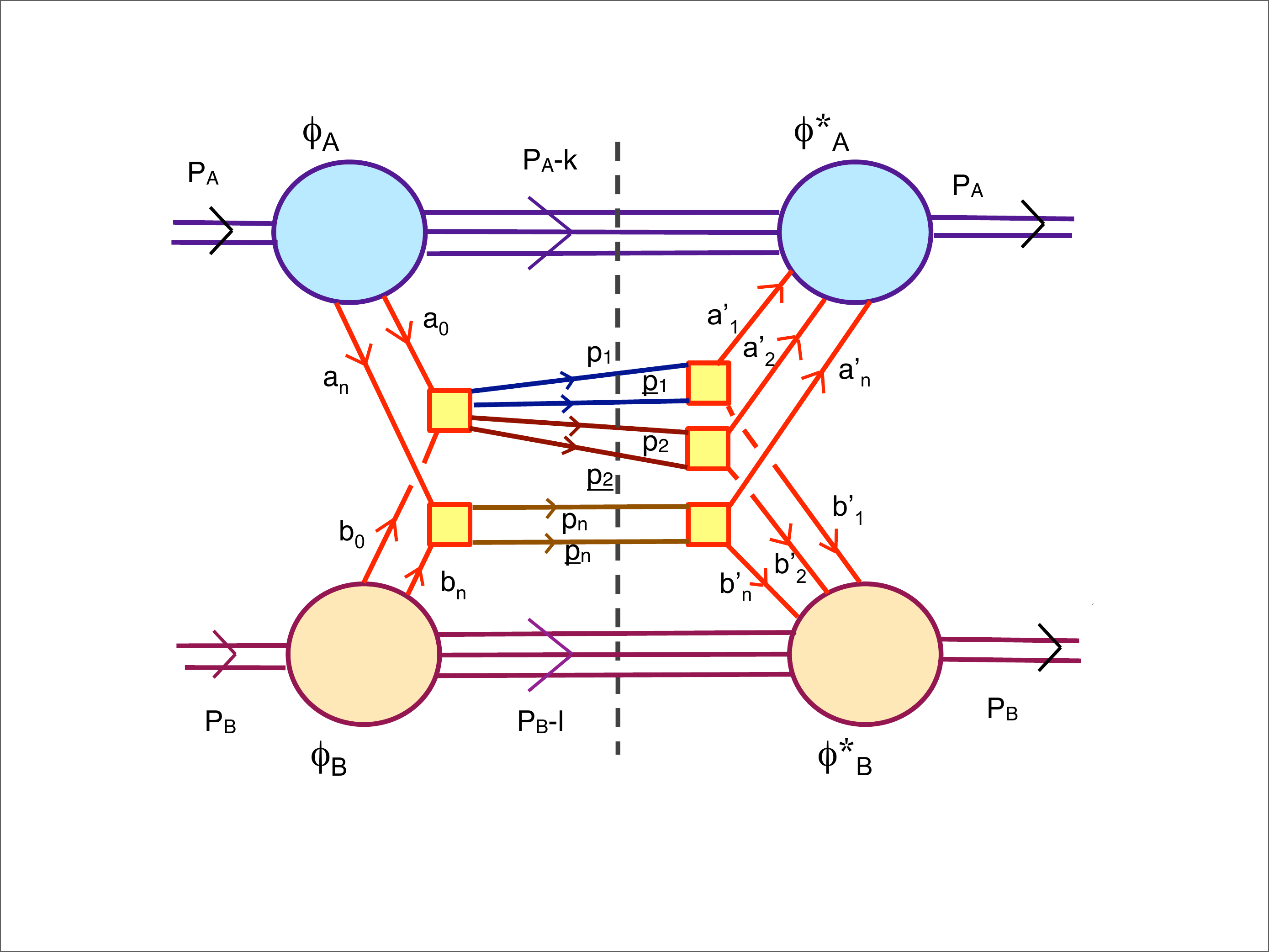}
\vspace{0cm}
\caption{Interference term}
\label{fig:interference}
\end{figure}

\noindent
In this case the independent variables are $3n-2$:

\begin{eqnarray}
\begin{cases}
P_i&n-\text{variables},\\
\delta_i&(n-2)-\text{variables},\\
\delta_i'&(n-1)-\text{variables},\\
k\quad\text{or}\quad l.
\end{cases}
\end{eqnarray}

\noindent
The longitudinal variables may be discussed as in the previous case. The obvious difference is in the resulting expression, which is no longer a modulus squared. The main point for the present considerations concerns the integrations on the transverse variables. Analogously to the diagonal case, the integrations on the transverse variables are conveniently discussed taking the Fourier transforms of the functions $\psi$:

\begin{eqnarray}
\psi_{A,n-1}\bigl(x_0^A,x_3^A&\dots& x_n^A; a_{0\perp},a_{3\perp}\dots a_{n\perp};(P_A-k)^2;j\bigr)\nonumber\\&=&\int\tilde\psi_{A,n-1}\bigl(x_0^A,x_3^A\dots x_n^A;\ s_0,s_3\dots s_n;(P_A-k)^2;j\bigr)\prod_{i=1}^n\esp{i a_{i\perp}\cdot s_i}\frac{d^2s_i}{2\pi}\nonumber\\
\psi_{B,n-1}\bigl(x_0^B,x_3^B&\dots& x_n^B;\ b_{0\perp},b_{3\perp}\dots b_{n\perp};(P_B-l)^2;j\bigr)\nonumber\\&=&\int\tilde\psi_{B,n-1}(x_0^B,x_3^B\dots x_n^B;\ \bar s_0,\bar s_3\dots\bar s_n;(P_B-l)^2;j\bigr)\prod_{i=1}^n\esp{i b_{i\perp}\cdot \bar s_i}\frac{d^2\bar s_i}{2\pi},
\end{eqnarray}

\noindent
where a label representing the number of interacting parton lines ($n-1$ in this case) has been introduced. The complex conjugate functions $\tilde\psi^*_n$ are the same as in the diagonal case. As discussed in the diagonal case, the large transverse momenta exchanged in each elementary collision localize the interactions in transverse space regions much smaller as compared to the hadron radius. While the discussion of the transverse variables may be done following the lines of the diagonal case, the treatment may be simplified by neglecting since start the distances between the interacting partons $r_i,\ r_i'$ in comparison with the parton coordinates $s_i,\ s_i',\ \bar s_i,\ \bar s_i'$. Introducing the variable $q_0=a_0-b_0$ one has

\begin{eqnarray}
\begin{cases}
a_0=\frac{1}{2}(P_1+P_2+q_0)\\
b_0=\frac{1}{2}(P_1+P_2-q_0).
\end{cases}
\end{eqnarray}

\noindent
The integration on $q_{0\perp}$ gives

\begin{eqnarray}
\int dq_{0\perp}&\to&\delta(s_0-\bar s_0+\beta),
\end{eqnarray}

\noindent
the integrations on $q_1',\ q_2'$ (previously defined) give

\begin{eqnarray}
\int dq_{1\perp}'&\to&\delta(s_1'-\bar s_1'+\beta)\nonumber\\
\int dq_{2\perp}'&\to&\delta(s_2'-\bar s_2'+\beta)
\end{eqnarray}

\noindent
and all other integrations on $q_i,\ q_i'$ are the same as in the diagonal case. At this stage the functions $\tilde\psi$ depend on the transverse variables as follows (the dependence on the fractional momenta and on the invariant mass of the remnants of the hadron is implicit)

\begin{eqnarray}
\tilde\psi_{A,n-1}(s_0,s_3\dots)\tilde\psi_{A,n}^*(s_1',s_2',s_3'\dots)\tilde\psi_{B,n-1}(s_0-\beta,s_3-\beta\dots)\tilde\psi_{B,n}^*(s_1'-\beta,s_2'-\beta,s_3'-\beta\dots).\nonumber\\
\end{eqnarray}

\noindent
One may now perform the integrations on $P_{1\perp},\ P_{2\perp}$. The result is

\begin{eqnarray}
\int dP_{1\perp}&\to&\delta(s_0+\bar s_0-s_1'-\bar s_1')\nonumber\\
\int dP_{2\perp}&\to&\delta(s_0+\bar s_0-s_2'-\bar s_2'),\nonumber\\
\end{eqnarray}

\noindent
which is equivalent to $\delta(s_0-s_1')\delta(s_0-s_2')$. All other integrations on $P_{i\perp}$ give the same result as in the diagonal case. One hence obtains that two transverse variables coincide in $\tilde\psi_{A,n}^*$ and $\tilde\psi_{B,n}^*$ and the cross section is proportional to the integral

\begin{eqnarray}
\int&&\tilde\psi_{A,n-1}(s_0,s_3\dots s_n)\tilde\psi_{A,n}^*(s_0,s_0,s_3\dots s_n)\nonumber\\
\times&&\tilde\psi_{B,n-1}(s_0-\beta,s_3-\beta\dots s_n-\beta)\tilde\psi_{B,n}^*(s_0-\beta,s_0-\beta,s_3-\beta\dots s_n-\beta)ds_0d\beta\prod_{i=3}^nds_i\nonumber\\
\end{eqnarray}

\noindent
Analogously to the diagonal case, $s_0,s_3\dots s_n$ represent the transverse coordinates of the positions of the interaction regions. One may hence conclude that, in the interference term between a $n$ and a $n-1$ collisions amplitude, the hard component of the interaction is localized in $n-1$ points in transverse space.

When the number of hard collisions is the same in both sides, a non-diagonal contribution may be obtained by linking, across the cut in Fig.1, two different collision amplitudes through the produced large $p_t$ final states. In such a case one obtains that the positions of the two hard interactions are localized within the same region, with size of ${\cal O} (r_{\perp}^2)$, which implies that the number of integrations in the transverse coordinates $s_{i\perp}$ is reduced by one unit with respect to the diagonal case. Also in this case the interference term hence corresponds to a case where the hard interaction is localized in $n-1$ points in transverse space. Rather obviously further crossings of the hard ($p,\ \bar p$) lines would further reduce the number of transverse integrations and hence the number of points in transverse space where the hard component of the interaction is localized.

\section{Concluding discussion}

A given multi-partons final state may be produced by interactions involving different numbers of partons in the initial state and the cross section, resulting from the coherent sum of all different terms, is expressed by a sum of diagonal and off diagonal contributions. As shown in Sec.2, the diagonal contribution, corresponding to a term with $n$ partons in the initial state, is given by the incoherent super-positions of $n$ disconnected parton interactions, localized in $n$ different points in transverse space. As shown in Sec.3, the hard component of the interaction corresponding to off diagonal contributions is disconnected and localized in no more than $n-1$ points in transverse space.

One may hence argue that interference terms do not represent corrections to the $n$-partons scattering inclusive cross section. They rather correct the $n-1$-partons (or less) scattering inclusive cross section.

Partons are in fact localized in the hadron by the momenta exchanged in the interaction. When partons are localized inside {\it non overlapping regions}, much smaller as compared to the hadron size, they are only connected one with another through soft exchanges and the picture of independent parallel collisions described in section 2 is a meaningful one. If, on the contrary, partons are localized by the interaction inside {\it overlapping regions}, much smaller as compared to the hadron size, they are allowed to interact by exchanging momenta of the size of their virtuality, which implies that the evaluation of the interference term, as discussed in section 3, is no longer adequate.

In Fig.3 we show an interference diagram between a single collision amplitude, where two partons interact at tree level producing four large $p_t$ partons, and an amplitude, where two partons, generated by the same short distance quantum fluctuation in the hadron $B$, interact with two partons of the hadron A, producing two pairs of large $p_t$ partons.

\begin{figure}[htp]
\centering
\includegraphics[width=13cm]{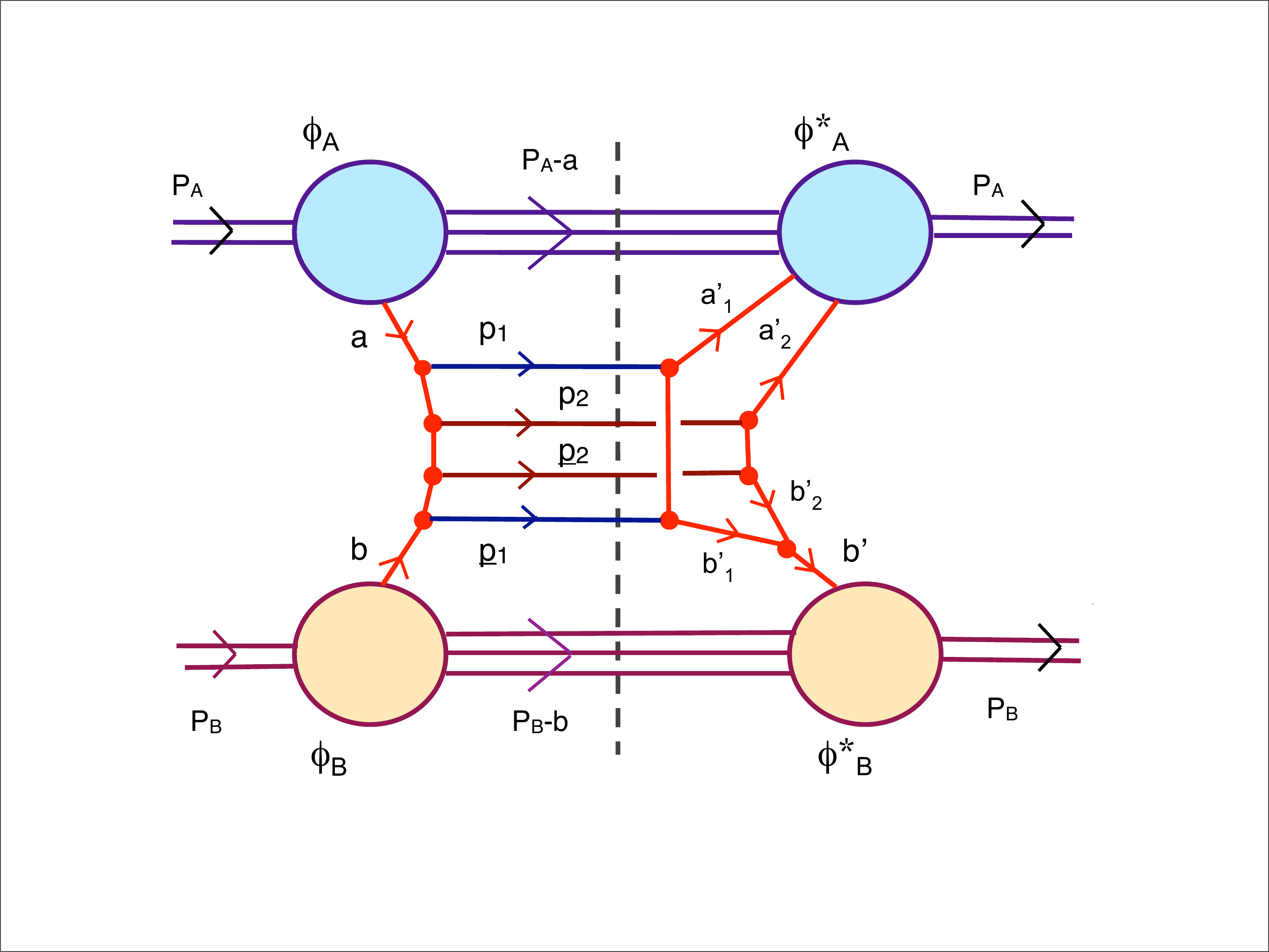}
\vspace{0cm}
\caption{A particular case of interference diagram}
\label{fig:interference between terms at different orders in g}
\end{figure}

\noindent
As it appears looking at Fig.3, because of the localization of the hard component of the interaction, the problem of interference is strictly linked to the problem to evaluate the single scattering amplitude, at higher orders in the coupling constant and including higher twists in the hadron structure.

One may hence conclude that a convenient way to distinguish the different terms in a MPI process is by their different topologies:

As a consequence of the different scales involved in the interaction, namely the hadron size and the large momenta exchanged, the structure of the hard component may be disconnected, with the different hard parts linked only through soft exchanges. The disconnected parts of the hard interaction are localized in different regions in transverse space, inside the overlap of the matter distribution of the two interacting hadrons. The different MPI terms are to be understood as the contributions to the final state due to the different disconnected parts of the hard component of the interaction.

In the instance of a single scattering the hard component is wholly connected and hence localized inside a single region in transverse space. In the simplest case, a single interaction is well described by the simple QCD-parton model recipe, with the parton interaction cross section evaluated at the lowest order in the coupling constant and convoluted with the parton distributions. In other cases, one may need to take into account higher order terms in the coupling constant to evaluate the partonic cross section or/and to include higher twist terms in the parton distributions. The effects of higher order terms and of higher twists may be more important when dealing with final states with several large $p_t$ partons.

A multiple interaction, on the other hand, has to be understood as a process where the hard part of the interaction is disconnected and localized in a number of different regions in transverse space. In the simplest case, in each different region the interaction may be evaluated at the lowest order in the coupling constant.

The main observation in the present note is that when MPI are understood in the topological sense described here above, different MPI terms, corresponding to different localizations in transverse space, namely to different topologies, do not interfere and the final cross section is obtained simply by the superposition of the cross sections, due to the contributions of the different topologies of the hard component of the interaction.

The topological feature represents on the other hand also the property which allows to recognize the contribution to the final state due to MPI. In each single parton collision all transverse momenta need to balance and a MPI process contributes to the cross section generating different groups of final state partons where the large transverse momenta are compensated separately. The compensation of transverse momenta within different subsets of large $p_t$ final state partons was in fact the feature which allowed the experimental identification and study of double parton collisions[13-16]. Interestingly, in a recent study of MPI at LHC energies it was shown that, requiring the compensation of transverse momenta within different subsets of large $p_t$ final state partons, one my expect to be able to isolate contributions due to triple parton collisions also in channels with a relatively low cross sections and with relatively large transverse momenta\cite{Maina:2009vx}.

\end{document}